\documentclass[nocopyrightspace, preprint]{sigplanconf}

\usepackage[utf8]{inputenc}
\usepackage[english]{babel}

\usepackage{amsthm}
\usepackage{todonotes}
\usepackage{amsmath}
\usepackage{amssymb}
\usepackage{url}
\usepackage{listings}
\usepackage{balance}
\usepackage{epigraph}
\usepackage{fixltx2e}
\usepackage{graphicx}
\usepackage{algorithm}
\usepackage{algpseudocode}
\usepackage{caption}
\usepackage{subcaption}

\theoremstyle{definition}

\theoremstyle{remark}

\begin{document}

\title{Declarative, Secure, Convergent Edge Computation}

\authorinfo{Christopher Meiklejohn}
           {}
           {christopher.meiklejohn@gmail.com}
\maketitle


\begin{abstract}
Eventual consistency is a more natural model than strong consistency for a distributed system, since it is closer to the underlying physical reality.  Therefore, we propose that it is important to find a programming model that is both congenial to developers and supports eventual consistency.  In particular, we consider that a crucial test for such a model is that it should support edge computation in a both natural and secure way.  We present a preliminary work report with an initial solution, called Lasp, which resembles a concurrent functional language while naturally supporting an eventually consistent coordination-free distribution model.
\end{abstract}

\section{Introduction}
Many of today's mobile applications and sensor networks are being built using a traditional client-server model, which places the data center at the center of computation.  In this model, clients generate data at the edge, usually a set of immutable events, that is then transmitted to a data center for processing by an application provider.  In the event that clients are disconnected and unable to transmit this data, two approaches have been used in practice: generated events can be buffered and transmitted once connectivity is restored or the client can prohibit the generation of events at the device by restricting application use.

This design is desirable to application developers because it pairs a familiar programming paradigm with an efficient system for executing computations.  This allows developers to compute with data that is centrally located, eliminating the inherent complexity required for computations that occur across a dynamic number of occasionally disconnected clients with varying network latencies.

While widely deployed in practice, this design is inherently limited as a result of both the limited storage and power capacity on edge devices.  In an ideal design, clients can operate with replicated, shared state and perform local computations: this design results in less overhead in state transmission and allows devices to make local decisions while disconnected from the network.

As an example, consider the case of an edge device responsible for monitoring the temperature of a refrigerator in a hospital.  One type of event this device might generate is an alert if the temperature becomes too warm.  While it is possible to buffer these events while the device is disconnected, it is preferable to enable the device to make local decisions for the sake of timeliness. 

\section{``Building On Quicksand''}
When moving to a model of computation with replicated, shared state, practitioners have typically treated a database as the ``source of truth'' for its data.  In this model, clients ask for a copy of some state, perform local mutations on that state, and then attempt to write this updated state back to the server.

When there are multiple clients in the system, read and write operations are interleaved amongst the clients.  As a result of this , multiple clients can attempt to concurrently modify the same state.  In a centralized approach with a single server, this is traditionally handled by a transaction protocol that only allows one of the concurrent operations to succeed.

However, when we move to a model where there are multiple replicas of some shared state, typically to reduce latency and increase fault-tolerance, it becomes increasingly difficult to enforce a total order without reducing availability of the system.~\cite{gilbert2002brewer}  Put in other words, to enforce a total order over concurrent modifications to replicated, shared state, synchronization is required for operation to complete; in the event that some nodes are not reachable due to network connectivity problems, the system can no longer make progress.  This issue is exacerbated as mobile and sensor network applications are deployed given periods without connectivity may cause more operations to appear as if they were concurrent.

As an alternative, several ``eventually consistent'' database systems have been developed. These systems typically offer weaker consistency guarantees and strive for high-availability and fault-tolerance.   Arguably the most famous of these systems, Amazon's Dynamo, only guaranteed that all updates would eventually be delivered to all replicas; without any ordering guarantees for events in the system, some concurrent modifications performed at different replicas would ultimately conflict and need to be resolved by the application developer.~\cite{decandia2007dynamo}

\section{``Single System Illusion''}
We posit that the ``eventually consistent'' view of the world is more compatible with these new types of applications, and more correctly models interactions between entities in the physical world.

For example, we can imagine a design where clients own the canonical copy of their data: this is an inversion of the traditional database model.  This data is locally mutated by clients and shared to other clients in the system: when other clients in this system mutate this state, this is represented as new data that is causally related to the original data; under concurrent modifications, these changes should be mergeable.  This data is disseminated in a peer-to-peer manner instead of synchronizing with a centralized server.

Computations, or instructions for deriving a value from this data, can also be disseminated between members of the system.  Computations reflect the completeness of their inputs: as more information, and more up-to-date information is provided, these computations should be able to be incrementally maintained.  Therefore, computations reflect derived data based on a partial view of the system.  It is important to note that this derived data can be disseminated without necessarily also supplying the source information: for instance, you might know that the Earth travels around the Sun without knowing \textit{why}.

Computations additionally compute causality, or a notion of provenance.~\cite{zaharia2012resilient, green2007provenance}  The result of a computation contains a record of the inputs that produced an output.  This allows the results of computations to be partially ordered, comparable, and mergeable.  Given computations are first class, they themselves can be causally related to other computations.

\section{``On The Road To Find Out''}
We previously proposed an initial solution to the problem of large-scale distributed programming with minimal coordination, named Lasp.~\cite{meiklejohn2015lasp} Lasp uses declarative, functional programming techniques to deterministically compose Conflict-Free Replicated Data Types (CRDTs), which model sequential data structures that when distributed, guarantee convergence under concurrent mutations and out-of-order message delivery.  This gives applications developed in Lasp a strong convergence property: given replicated state that is concurrently edited and eventually communicated to every node in a distributed system regardless of ordering, distributed applications will converge to the correct result.~\cite{shapiro2011conflict}  Lasp's epidemic-based distributed runtime complements this model well: we can take advantage of an optimized dissemination protocol with no guarantees on message ordering when the programming model is tolerant to message reordering and replay.~\cite{meiklejohn2015selective}  However, while Lasp provides the basic building blocks for building distributed, convergent computations, there are still fundamental problems to solve.

One problem is causality.  Causality is necessary for the incremental maintenance and mergeability of computation results.  For example, if we have a replicated, shared set and we compute some function over that set, how can we efficiently represent the input to the computation in the output: this allows us to ensure results can be incrementally maintained and merged with replicated copies of the same computation as the inputs to the computation change over time.  Lasp currently provides this functionality for a limited set of functional and set-theoretic operations over CRDTs.  However, the question remains as to whether these mechanisms can be extended to arbitrary higher-order programming.

Another problem is security.  If we move all computation to the edge, how can we write distributed computations where each member of the system can incrementally contribute to a final result in a way where the individual user's values are not exposed?  Returning to the causality example in the previous paragraph, how can we securely compare causality information to determine if values are stale?  Given the semilattice properties of CRDTs, we wonder if there is a way to leverage order-preserving encryption: in our refrigerator example, is there a way where individual units could alert on temperature conditions securely, without exposing their actual temperatures to other units in the system.~\cite{kolesnikov2012limits}

Finally, the problem of expressiveness.  In this model, if we can make very few guarantees on when, and in what order, events will be seen by all members of the system, does this restrict the space of possible programs that can be expressed within this model?  In what ways do the requirements of CRDTs, because data structures must be associative, commutative, and idempotent, restrict what types of abstractions a user can build within this programming model?

\section{Related Work}
Applications today that perform data processing from devices at the edge use a MapReduce-style programming model over immutable data~\cite{dean2008mapreduce} and subsequent optimizations for efficient, fault-tolerant processing over streams.~\cite{zaharia2012resilient} These solutions are appealing to the application developer: they present systems for performing efficient computation in a now-familiar programming paradigm.

Alternative techniques have been presented by academia that focus on moving computation to the edge to alleviate the need for transmission of the entire data set.  Directed \cite{intanagonwiwat2000directed} and digest~\cite{zhao2003computing} diffusion presented efficient, fault-tolerant approaches for dissemination of computations and their results.  However, these systems do not expose a general programming model.

Declarative approaches such as Tiny AGgregation, \cite{madden2002tag} have been proposed for data collection and aggregation across sensor networks.  However, these approaches have presented abstractions that are not for general programming as they are specific to the details of aggregation in sensor networks.  Finally, declarative approaches~\cite{lorenz2015separation} have also been proposed for computation in large-scale peer-to-peer systems where clients own their own data.  However, in an effort to make their language Turing complete, they relied on the programmer explicitly encoding the details around how statements would be evaluated, to ensure termination. 

\section{Conclusion}
As we move towards application designs that operate over a large amount of client generated data in a privacy-conscious world, is it possible to define a natural way of computing securely at the edge?

\acks
Thanks to Peter Alvaro and Peter Van Roy.  This work has been partially funded by the SyncFree EU/FP7 Project (n\textsuperscript{o} 609551).

\balance

\bibliographystyle{abbrvnat}
\bibliography{obt-2016}

\begin{thebibliography}{13}
\providecommand{\natexlab}[1]{#1}
\providecommand{\url}[1]{\texttt{#1}}
\expandafter\ifx\csname urlstyle\endcsname\relax
  \providecommand{\doi}[1]{doi: #1}\else
  \providecommand{\doi}{doi: \begingroup \urlstyle{rm}\Url}\fi

\bibitem[Dean and Ghemawat(2008)]{dean2008mapreduce}
J.~Dean and S.~Ghemawat.
\newblock Mapreduce: simplified data processing on large clusters.
\newblock \emph{Communications of the ACM}, 51\penalty0 (1):\penalty0 107--113,
  2008.

\bibitem[DeCandia et~al.(2007)DeCandia, Hastorun, Jampani, Kakulapati,
  Lakshman, Pilchin, Sivasubramanian, Vosshall, and Vogels]{decandia2007dynamo}
G.~DeCandia, D.~Hastorun, M.~Jampani, G.~Kakulapati, A.~Lakshman, A.~Pilchin,
  S.~Sivasubramanian, P.~Vosshall, and W.~Vogels.
\newblock {Dynamo: {Amazon}'s highly available key-value store}.
\newblock In \emph{ACM SIGOPS Operating Systems Review}, volume~41, pages
  205--220. ACM, 2007.

\bibitem[Gilbert and Lynch(2002)]{gilbert2002brewer}
S.~Gilbert and N.~Lynch.
\newblock Brewer's conjecture and the feasibility of consistent, available,
  partition-tolerant web services.
\newblock \emph{ACM SIGACT News}, 33\penalty0 (2):\penalty0 51--59, 2002.

\bibitem[Green et~al.(2007)Green, Karvounarakis, and
  Tannen]{green2007provenance}
T.~J. Green, G.~Karvounarakis, and V.~Tannen.
\newblock Provenance semirings.
\newblock In \emph{Proceedings of the twenty-sixth ACM SIGMOD-SIGACT-SIGART
  symposium on Principles of database systems}, pages 31--40. ACM, 2007.

\bibitem[Intanagonwiwat et~al.(2000)Intanagonwiwat, Govindan, and
  Estrin]{intanagonwiwat2000directed}
C.~Intanagonwiwat, R.~Govindan, and D.~Estrin.
\newblock Directed diffusion: a scalable and robust communication paradigm for
  sensor networks.
\newblock In \emph{Proceedings of the 6th annual international conference on
  Mobile computing and networking}, pages 56--67. ACM, 2000.

\bibitem[Kolesnikov and Shikfa(2012)]{kolesnikov2012limits}
V.~Kolesnikov and A.~Shikfa.
\newblock On the limits of privacy provided by order-preserving encryption.
\newblock \emph{Bell Labs Technical Journal}, 17\penalty0 (3):\penalty0
  135--146, 2012.

\bibitem[Lorenz and Rosenan(2015)]{lorenz2015separation}
D.~H. Lorenz and B.~Rosenan.
\newblock Separation of powers in the cloud: where applications and users
  become peers.
\newblock In \emph{2015 ACM International Symposium on New Ideas, New
  Paradigms, and Reflections on Programming and Software (Onward!)}, pages
  76--89. ACM, 2015.

\bibitem[Madden et~al.(2002)Madden, Franklin, Hellerstein, and
  Hong]{madden2002tag}
S.~Madden, M.~J. Franklin, J.~M. Hellerstein, and W.~Hong.
\newblock Tag: A tiny aggregation service for ad-hoc sensor networks.
\newblock \emph{ACM SIGOPS Operating Systems Review}, 36\penalty0
  (SI):\penalty0 131--146, 2002.

\bibitem[Meiklejohn and Van~Roy(2015{\natexlab{a}})]{meiklejohn2015lasp}
C.~Meiklejohn and P.~Van~Roy.
\newblock {Lasp: A Language for Distributed, Coordination-Free Programming}.
\newblock In \emph{Proceedings of the 17th International Symposium on
  Principles and Practice of Declarative Programming}, pages 184--195. ACM,
  2015{\natexlab{a}}.

\bibitem[Meiklejohn and Van~Roy(2015{\natexlab{b}})]{meiklejohn2015selective}
C.~Meiklejohn and P.~Van~Roy.
\newblock {Selective Hearing: An Approach to Distributed, Eventually Consistent
  Edge Computation}.
\newblock In \emph{Workshop on Planetary-Scale Distributed Systems collocated
  with SRDS 2015}. IEEE, 2015{\natexlab{b}}.

\bibitem[Shapiro et~al.(2011)Shapiro, Pregui{\c{c}}a, Baquero, and
  Zawirski]{shapiro2011conflict}
M.~Shapiro, N.~Pregui{\c{c}}a, C.~Baquero, and M.~Zawirski.
\newblock Conflict-free replicated data types.
\newblock In \emph{Stabilization, Safety, and Security of Distributed Systems},
  pages 386--400. Springer, 2011.

\bibitem[Zaharia et~al.(2012)Zaharia, Chowdhury, Das, Dave, Ma, McCauley,
  Franklin, Shenker, and Stoica]{zaharia2012resilient}
M.~Zaharia, M.~Chowdhury, T.~Das, A.~Dave, J.~Ma, M.~McCauley, M.~J. Franklin,
  S.~Shenker, and I.~Stoica.
\newblock Resilient distributed datasets: A fault-tolerant abstraction for
  in-memory cluster computing.
\newblock In \emph{Proceedings of the 9th USENIX conference on Networked
  Systems Design and Implementation}, pages 2--2. USENIX Association, 2012.

\bibitem[Zhao et~al.(2003)Zhao, Govindan, and Estrin]{zhao2003computing}
J.~Zhao, R.~Govindan, and D.~Estrin.
\newblock Computing aggregates for monitoring wireless sensor networks.
\newblock In \emph{Sensor Network Protocols and Applications, 2003. Proceedings
  of the First IEEE. 2003 IEEE International Workshop on}, pages 139--148.
  IEEE, 2003.

\end{thebibliography}

\end{document}